\documentclass[twocolumn,showpacs,preprintnumbers,amsmath,nofootinbib,amssymb]{revtex4}
\input psfig.sty
\def \be {\begin{equation}}
\def \ee {\end{equation}}
\def \bea {\begin{eqnarray}}
\def \eea {\end{eqnarray}}

\usepackage{graphicx}% Include figure files
\usepackage{dcolumn}% Align table columns on decimal point
\usepackage{bm}% bold mathneralized

\begin{document}

\title{CMB and LSS constraints on a single-field model of  inflation}

\author{F. C. Carvalho$^1${\footnote{fabiocc@on.br}}}

\author{J. S. Alcaniz$^1${\footnote{alcaniz@on.br}}}

\author{J. A. S. Lima$^2${\footnote{limajas@astro.iag.usp.br}}}

\author{R. Silva$^3${\footnote{raimundosilva@uern.br}}}

\affiliation{$^1$Observat\'orio Nacional, 20921-400 Rio de Janeiro - RJ, Brasil}

\affiliation{$^2$Universidade de S\~ao Paulo, 05508-900 S\~ao Paulo, SP, Brasil}

\affiliation{$^3$Universidade do Estado do Rio Grande do Norte, 59610-210, Mossor\'o, RN, Brasil}

\date{\today}

\begin{abstract}
A new inflationary  scenario whose exponential potential $V(\Phi)$ has a quadratic dependence on the field $\Phi$ in addition to the standard linear term is confronted with the tree-year observations of the Wilkinson-Microwave Anisotropy Probe and the Sloan Digital Sky Survey data.  The number of e-folds ($N$), the ratio of tensor-to-scalar perturbations ($r$), the spectral scalar index of the primordial power spectrum ($n_s$) and its running ($dn_s/d\ln k$) depend on the dimensionless  parameter $\alpha$ multiplying the quadratic term in the potential. In the limit $\alpha \rightarrow 0$ all the results of the standard exponential potential are fully recovered. For values of $\alpha \neq  0$, we find that the model predictions are in good agreement with the current observations of the Cosmic Microwave Background (CMB) anisotropies and Large-Scale Structure (LSS) in the Universe.

\end{abstract}

\maketitle

\section{Introduction}

Since the detection of the Cosmic Microwave Background (CMB) anisotropies by the {\rm COBE} satellite in 1992 \cite{cobe}, great improvements in the quality of the CMB data has been achieved, mainly from very recent baloon and satellite experiments, such as the BOOMERANG \cite{boomerang} and the Wilkinson Microwave Anisotropy Probe ({\rm WMAP}) \cite{wmap,wmap3}. In what concerns the possible implications on the inflationary epoch, for instance, the current three-year WMAP (WMAP3) data seemed, at a first moment, to have precision enough to discriminate between some single-field inflationary models. In fact, soon after the WMAP3 data release, some authors arrived to the conclusion that quartic chaotic inflationary scenarios of the form $V(\Phi)\sim\lambda\Phi^4$ were ruled out, while quadratic chaotic inflationary models with potential, $V(\Phi)\sim m^2\Phi^2$, agreed only marginally with the observational data \cite{wmap3,easther,lyth}.  The original WMAP3 parameter-estimation analysis also pointed to a spectral index smaller ($n_s \simeq 0.95$)  than the Zel'dovich spectrum of density perturbations, for which adiabatic perturbations have a scale-invariant spectral index ($n_s=1$). Although compatible with some theoretical predictions \cite{mukanov}, the latter conclusion was rediscussed by a recent analysis of the WMAP3 data from a more sophisticated statistical approach of model selection and systematic effects, leading to values of $n_s$ still compatible with the Zel'dovich prediction \cite{melchi}.

The WMAP3 data also place an upper limit on the  tensor-to-scalar ratio, i.e., $r<0.55$ (at 95.4\% c.l.), whereas a joint analysis involving the WMAP3 data and the large-scale power spectrum of luminous red galaxies in the Sloan Digital Sky Survey (SDSS)  provides $r < 0.33$ (also at 95.4\% c.l.) \cite{tegmark}. In light of all these observational results, a number of authors have tested the viability of different types of inflationary models (see, e.g., \cite{easther,lyth,scenarios,beta,b1}). As an example, very recently,  the authors of Ref. \cite{b1} revived a interesting phenomenological model with a simple slowly-rolling scalar field that, in the light of the WMAP3 data, does not present a pure de Sitter inflationary expansion, but  produce a Zel'dovich spectrum, i.e., $n_s=1$.

Given the current availability of high precision cosmological data and, as consequence, the real possibility of truly ruling out some theoretical scenarios, it is timely to revive old inflationary models (as done in Ref.~\cite{b1}), as well as to investigate new ones. In this paper, motivated by a transient dark energy scenario recently proposed in Ref. \cite{prl}, we study a single, minimally-coupled scalar field model of inflation whose evolution is described by an exponential potential $V(\Phi)$ that has a quadratic dependence on the field $\Phi$ in addition to the standard linear term. Such a potential is obtained through a simple \emph{ansatz} and fully reproduces the Ratra-Peebles scenario studied in Ref. \cite{exp} (see also \cite{r1,r2}) in the limit of the dimensionless parameter $\alpha \rightarrow 0$. For all values of $\alpha \neq 0$, however, the potential is dominated by the quadratic contribution %present in the exponential function 
and admits a wider range of solutions than do conventional exponential potentials. 

In this context, our aim here is to test the viability of this new class of inflationary scenario in light of the current CMB and LSS data. In Sec. II we deduce the inflaton potential $V(\Phi)$ and discuss the basic features of the model. The slow-roll inflation driven by this potential along with some important observational quantities, such as the spectral index, its running, and the ratio of tensor-to-scalar perturbations, are discussed in Sec. III. We also confront our theoretical results with the most recent CMB and LSS observations, as analized in Refs.~\cite{wmap3,tegmark,melchi}. Finally, the main results of this paper are discussed and summarized in the Sec. IV.

\begin{figure}[t]
%\label(Vphi)
%\vspace{.2in}
\centerline{\psfig{figure=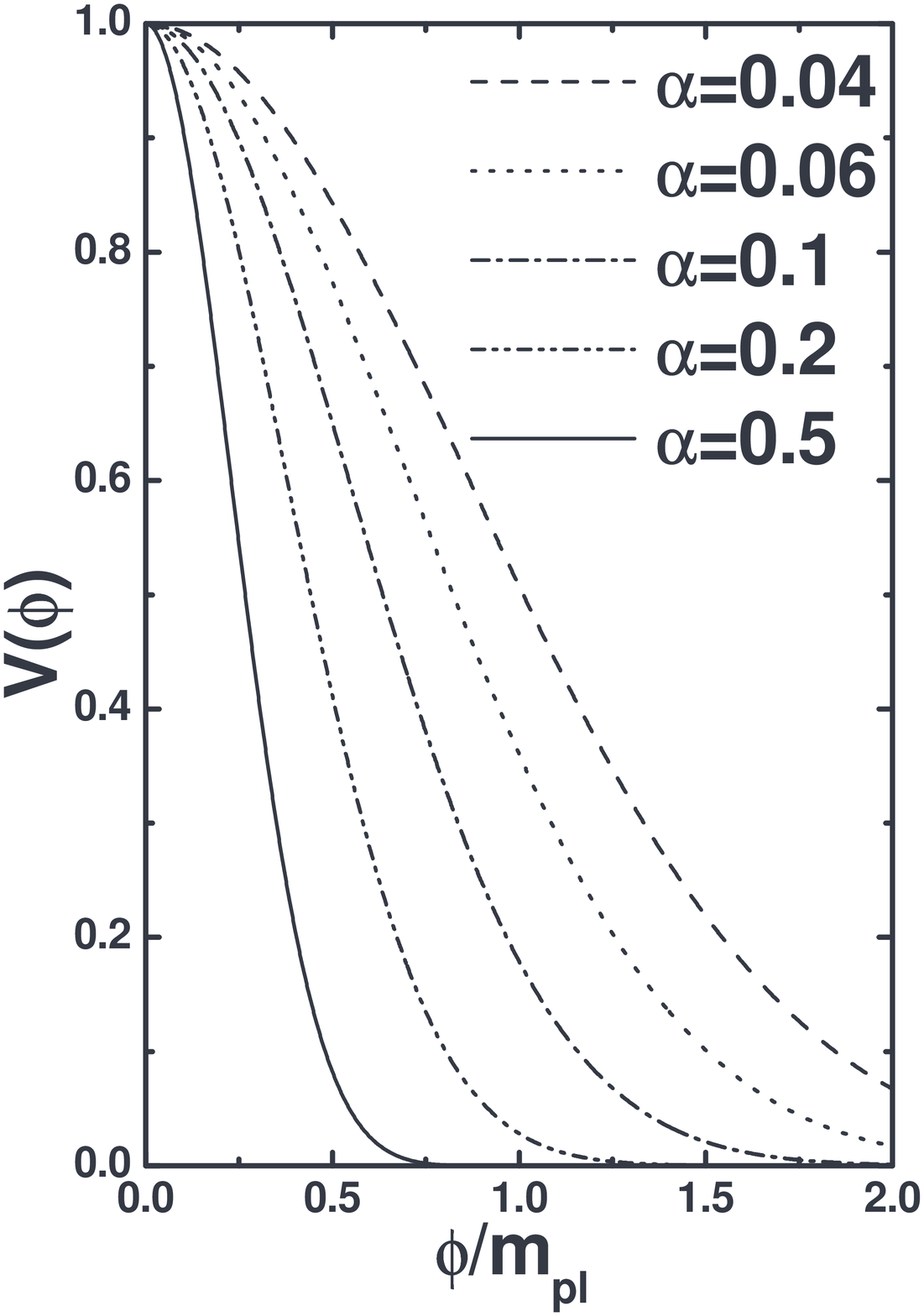,width=3.0truein,height=2.6truein}
\hskip 0.1in}
\caption{The potential $V(\Phi)$ as a function of the field
[Eq. (\ref{gpotential})] for some selected values of the parameter $\alpha$.}
\end{figure}

\section{Single-field model}
In what follows we assume that the Universe is nearly flat, as evidenced by the combination of the position of the first acoustic peak of the CMB power spectrum and the current value of the Hubble parameter \cite{wmap3}.  To begin with, let us consider a single scalar field model whose action is given by  
\be
\label{action} 
S=\frac{m^2_{\rm pl}}{16\pi}\int d^4 x \sqrt{-g}\left[R - \frac{1}{2}%g^{\mu\nu}
\partial^{\mu}\Phi\partial_{\mu}\Phi - V(\Phi)\right]\;.
\ee 
In the above expression,  $m_{\rm pl}\equiv G^{-1/2}\approx10^{19}GeV$ is the Planck mass and we have set the speed of light $c = 1$. 

For an inflaton-dominated universe, the Friedmann equation is written as
\begin{equation}
\label{b1}
H^2  = \left(\frac{\dot a}{a}\right)^2 = \frac{8 \pi}{3 m_{\rm pl}^2} \left[\frac{1}{2} \dot\Phi^2 + V\left(\Phi\right)\right]\;,
\end{equation}
where $a(t)$ is the cosmological scalar factor and dots denote derivatives with respect to time.  By combining Eq. (\ref{b1}) with the conservation equation for the $\Phi$ component, i.e.,  $\dot\rho_{\Phi}+3H(\rho_{\Phi}+p_{\Phi})=0$, we obtain
\be
\label{phi11} 
\frac{\partial\Phi}{\partial a} =\sqrt{-\frac{m_{\rm pl}^2}{8\pi
a}\frac{1}{\rho_{\Phi}}\frac{\partial\rho_{\Phi}}{\partial a}}\;,
\ee 
where $\rho_{\Phi}=\frac{1}{2}\dot\Phi^2+V(\Phi)$ and $p_{\Phi}=\frac{1}{2}\dot\Phi^2-V(\Phi)$ are, respectively, the inflaton energy density and pressure.

\begin{figure}[t]
%\label(Vphi)
%\vspace{.2in}
\centerline{\psfig{figure=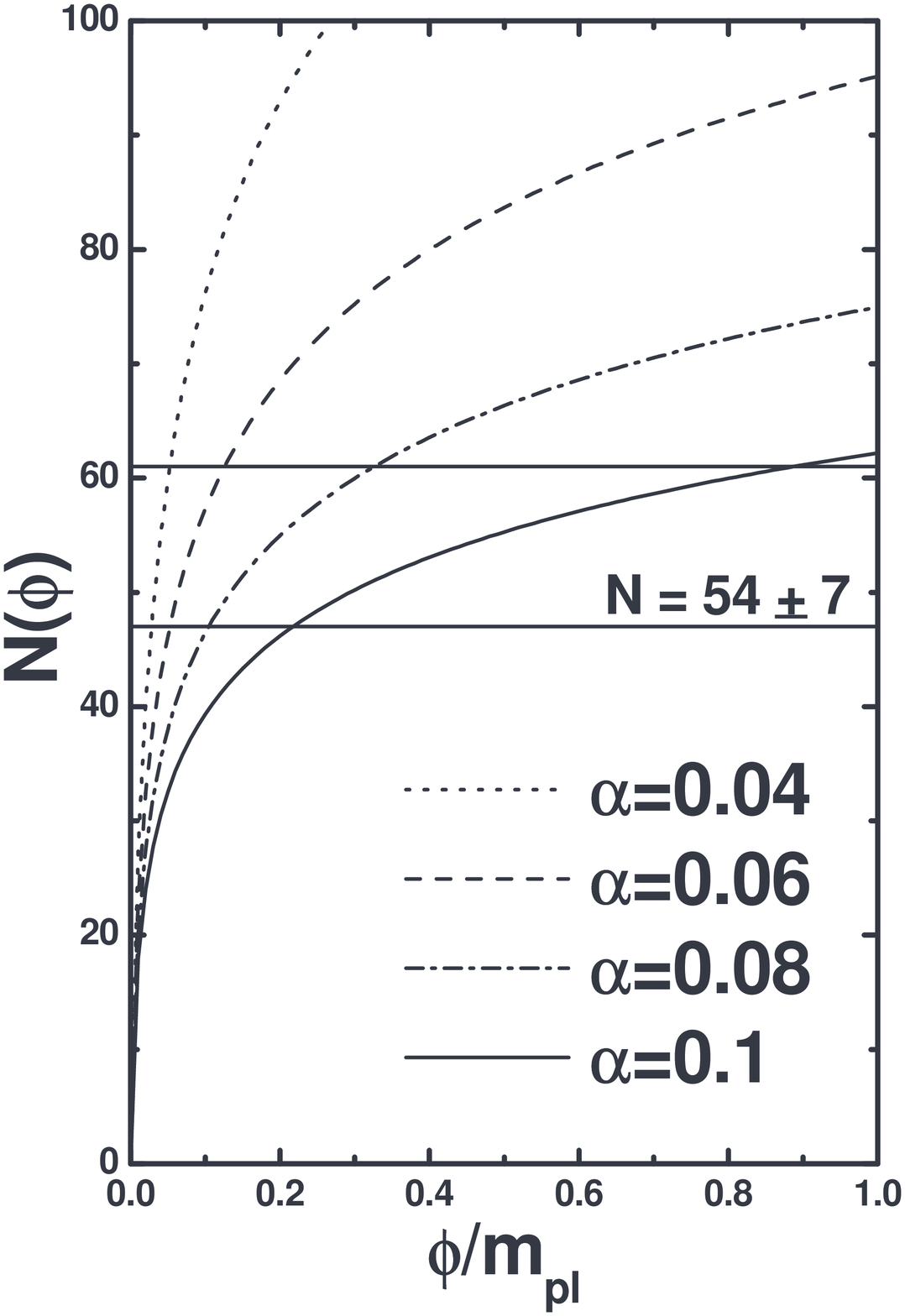,width=3.0truein,height=2.6truein}
\hskip 0.1in} 
\caption{The predicted number of {\sl e-folds} $N(\Phi)$ as a function of the field for some selected values of the parameter $\alpha$. The horizontal lines correspond to 1$\sigma$ limit on the number of {\sl e-folds} ($N = 54 \pm 7$) discussed in Ref.~\cite{e-fold}.}
\end{figure}

Following Ref. \cite{prl}, we adopt an \emph{ansatz} on the scale factor derivative of the energy density, i.e.,
\be
\label{ansatz}
\frac{1}{\rho_{\Phi}}\frac{\partial\rho_{\Phi}}{\partial
a}=-\frac{\lambda}{a^{1-2\alpha}}\;, 
\ee 
where $\alpha$ and $\lambda $ are positive parameters, and the factor 2 was introduced for mathematical convenience. From a direct combination of Eqs. (\ref{phi11}) and (\ref{ansatz}), the following expression for the scalar field is obtained 
\bea \label{phi1} 
\Phi(a)= \frac{1}{\sqrt{\sigma}}\ln_{1-\alpha}\left({a}\right)\;, 
\eea 
where  $\sigma = {8\pi/\lambda m_{\rm pl}^2}$ and the generalized logarithmic function $\ln_{1 - \xi}$, defined as $\ln_{1 - \xi}(x)\equiv{(x^{\xi}-1)/\xi}$, reduces to the ordinary logarithmic function in the limit $\xi \rightarrow 0$ \cite{abramowitz}. The potential $V(\Phi)$ for  the above scenario is easily derived by using the definitions of $\rho_{\Phi}$ and $p_{\Phi}$ and inverting\footnote{Note that the inversion of Eq. (\ref{phi1}) can be more directly obtained if one defines the generalized exponential function as $\exp_{1-\xi}(x)\equiv[1 + \xi{x}]^{1/\xi}$, which not only reduces to an ordinary exponential in the limit $\xi \rightarrow 0$ but also is the inverse function of the generalized logarithm ($\exp_{{1-\xi}}[{\ln_{1 -\xi}}(x)]=x$). Thus, the scale factor in terms of the field can be written as $a(\Phi)=\exp_{1-\alpha}[\sqrt{\sigma}\Phi]$ \cite{prl}.} Eq. (\ref{phi1}), i.e.,
\be
\label{gpotential} 
V(\Phi) = f(\alpha; \Phi) \exp\left[-\lambda \sqrt{\sigma}\left(\Phi +\frac{\alpha\sqrt{\sigma}}{2} \Phi^2 \right)  \right], 
\ee
where $f(\alpha; {\Phi}) \propto [1-\frac{\lambda}{6}(1+\alpha\sqrt{\sigma}{\Phi})^2]$. % and $\Phi = \Phi-\Phi_{\rm i}$. 
The most important aspect to be emphasized at this point is that in the limit $\alpha \rightarrow 0$ Eqs. (\ref{phi1}) and (\ref{gpotential}) fully reproduce the exponential potential studied by Ratra and Peebles in Ref. \cite{exp}, while $\forall$ $\alpha \neq 0$ the scenario described above represents a generalized model which admits a wider range of solutions. This means that all the physical observational quantities derived in the next section have the ordinary exponential case as a particular limit when $\alpha \rightarrow 0$. For the sake of completeness, in Fig.(1) we show the potential $V(\Phi)$ as a function of the field  for several values of the parameter $\alpha$ and a fixed value of $\lambda = 10^{-6}$ (see \cite{prl} for details).

\begin{figure}
%\label(Vphi)
\centerline{\psfig{figure=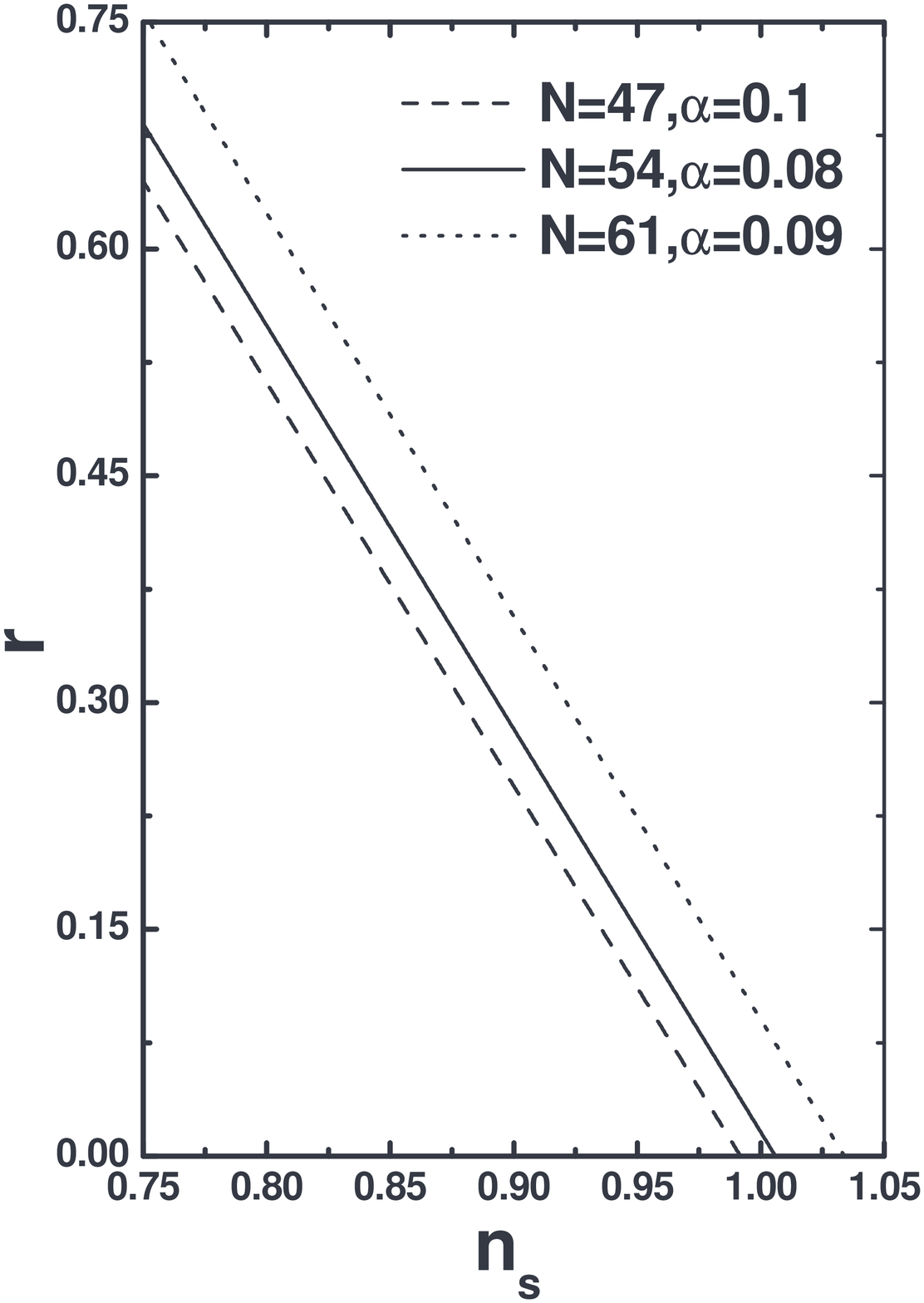,width=3.0truein,height=2.6truein}
\hskip 0.1in} 
\caption{The $n_s-r$ plane for some selected values of the parameter $\alpha$ to first-order in {\sl slow-roll} approximation. Note that, similarly to the intermediate inflationary model of Ref. \cite{b1}, it is possible to obtain a scale-invariant spectrum for nonvanishing values of $r$.}
\end{figure}

\section{{\sl Slow-roll} Inflation}

\subsection{{\sl Slow-roll} Parameters}

In this background, the energy conservation law for the field can be expressed as 
%\begin{equation}
$\ddot\Phi + 3 H \dot\Phi + V'\left(\Phi\right) = 0$, 
%\end{equation}
where primes denote derivative with respect to the field $\Phi$. In the so-called {\sl slow-roll} approximation, the evolution of the field is dominated by the drag from the cosmological expansion, so that $\ddot\Phi\approx0$ or, equivalently, $3H\dot \Phi+V'\simeq 0$. With these simplifications, the {\sl slow-roll} regime can be expressed in terms of the {\sl slow-roll} parameters  $\epsilon$ 
and $\eta$, i.e., \cite{book,lyth1}
\bea
\label{epsilon}
\epsilon = \frac{m^2_{\rm pl}}{16\pi}\left(\frac{V'}{ V}\right)^2
=\frac{\lambda}{2}\frac{[\lambda y^2 - 2(\alpha+3)]^2y^2}{(6-\lambda y^2)^2}\; ,
\eea
and
\bea	
\label{eta}
\eta = \frac{m^2_{\rm pl}}{ 8\pi}\frac{V''}{V}
=\frac{(5\alpha+6)\lambda y^2-\lambda^2 y^4 
- 2\alpha(\alpha+3)}{6-\lambda y^2}
\;,
\eea
where, for the sake of simplicity, we have introduced the variable $y=1+\alpha\sqrt{\sigma}\Phi$. Note that, in the limit $\alpha \rightarrow 0$, the above expressions reduce, respectively, to $\epsilon_{\alpha \rightarrow 0} = \frac{\lambda}{2}$ and $\eta_{\alpha \rightarrow 0} = \lambda$, as expected from conventional exponential potentials.

\begin{figure*}[t]
\centerline{
\psfig{figure=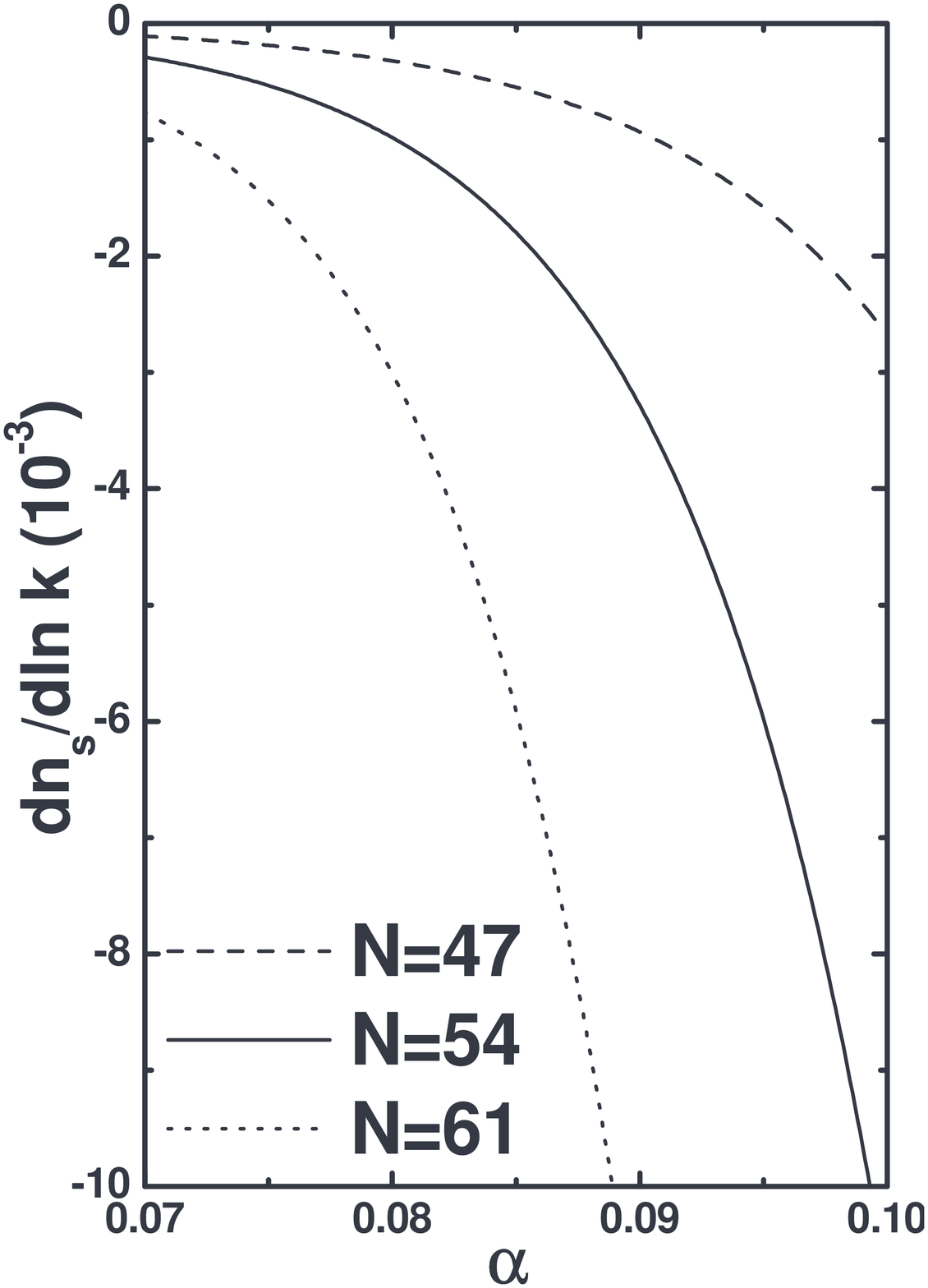,width=3.7truein,height=3.0truein,angle=0}
\psfig{figure=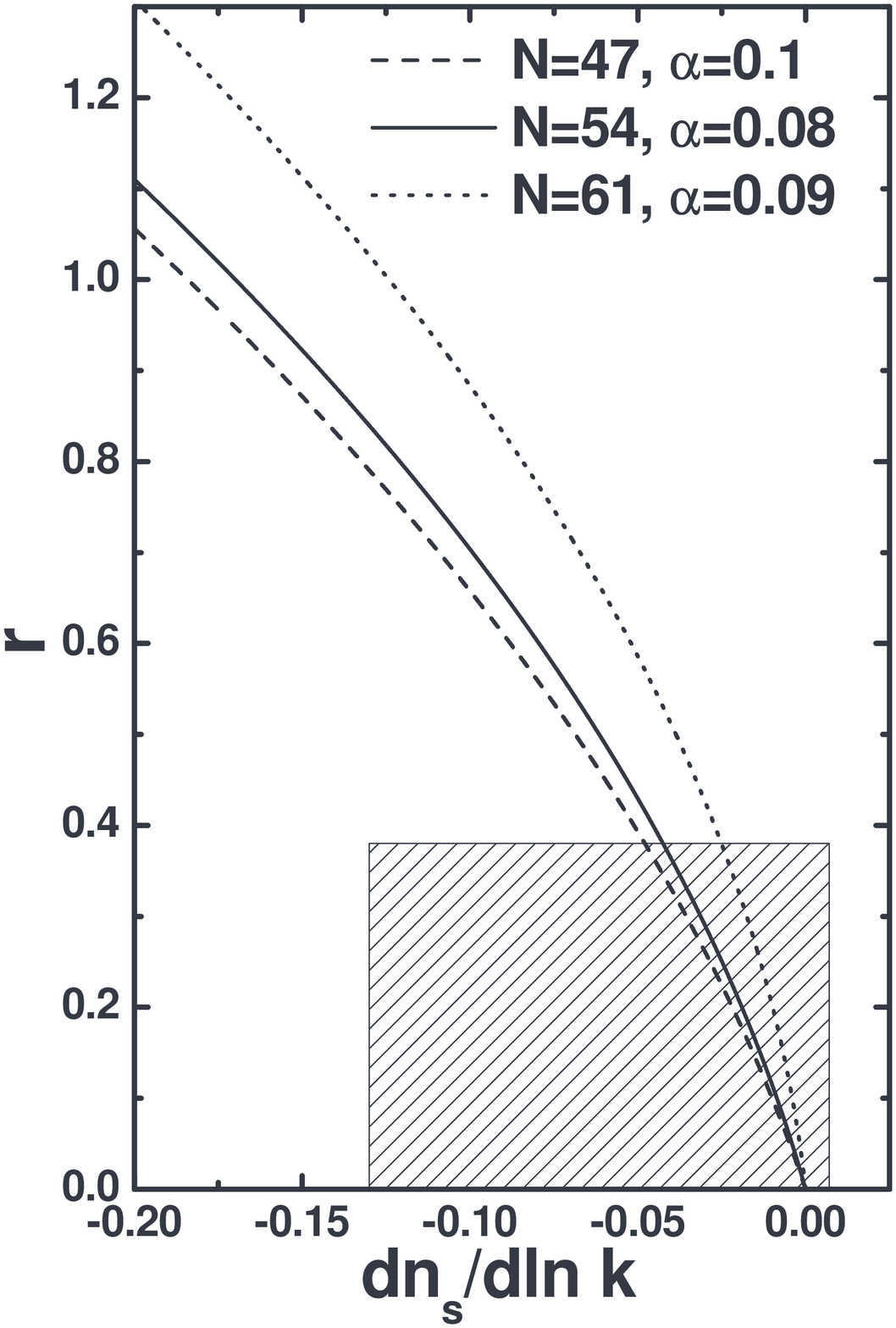,width=3.7truein,height=3.0truein,angle=0}
\hspace{0.1cm}}
\caption{{\bf{a)}} The $\alpha-dn_s/d \ln k$ plane for the number of {\sl e-folds} lying in the interval $N = 54 \pm 7$. Note that, as $\alpha \rightarrow 0$, $dn_s/d \ln k$ also $\rightarrow 0$, as expected from exponential potentials. The same conclusion on $dn_s/d \ln k$ also persists for negative values of the index $\alpha$. {\bf{b)}}  The $dn_s/d \ln k-r$ plane for $\alpha = 0.1, 0.09$ and 0.08. Here, the hachured region corresponds to $-0.13 \leq dn_s/d\ln k \leq 0.007$ and $r < 0.38$ (at 95.4\%), as given in Ref.~\cite{melchi}.}
\end{figure*}

For the above scenario, we can also compute the predicted number of  {\sl e-folds} by using Eq. (\ref{phi1}) and (\ref{epsilon}), i.e., $N = \int Hdt  = \ln\left[1+\alpha\sqrt{\sigma}\Phi_N\right]^{1/\alpha}$,
which reduces, in the limit $\alpha \rightarrow0$, to $N_{\alpha \rightarrow 0} \propto \Phi_{\rm N}$. The result of this calculation is shown in Fig. (2) as the $N - \Phi$ plane for some selected values of the index $\alpha$. The horizontal lines in the figure correspond to the 1$\sigma$ bound on the  number of  {\sl e-folds} discussed in  Ref. \cite{lyth}, i.e., $N = 54 \pm 7$. To test the viability of the inflationary scenario here discussed, in all the subsequent analyses we follow Ref. \cite{lyth} and adopt the interval $N = 54 \pm 7$. Without loss of generality, we also fix the value of the constant $\lambda$ at $\simeq 10^{-6}$.

\subsection{Spectral Index}

In order to confront our model with current observational results we first consider the spectral index, $n_s$, and the ratio of tensor-to-scalar perturbations, $r$. In terms of the {\sl slow-roll} parameters to first order, these quantities, defined as $n_s-1 = 2\eta- 6 \epsilon$ and $r=16\epsilon$, are now expressed as
\begin{widetext}
\bea
\label{power_s}
n_s-1 = -\frac{[\lambda y^2 - 2(\alpha+3)]^2y^2}{(6-\lambda y^2)^2}
[3\lambda+2(6-\lambda y^2)] + \frac{2\lambda y^2(\alpha-6)+4(\alpha+3)(\alpha+6)}{6-\lambda y^2},
\eea
\end{widetext}
and
\bea
\label{r}
r = 8\lambda\frac{[\lambda y^2 - 2(\alpha+3)]^2y^2}{(6-\lambda y^2)^2}\;.
\eea
As can be easily verified, in the limit $\alpha\rightarrow0$,  the above expressions reduce, respectively, to $(n_s-1)_{\alpha\rightarrow0}=-\lambda$ and $r_{\alpha\rightarrow0}=8\lambda$. For $r < 0.55$ (95.4\% c.l.), as given by current CMB data \cite{wmap3}, one obtains from Eq. (\ref{r}) $\epsilon < 0.03$, which is in agreement with the {\sl slow-roll} approximation discussed earlier and adopted in our analysis.  

Figure (3) shows  the $n_s - r$ plane, given by
\begin{equation}
\label{nsr}
r=\frac{8}{\gamma-3}(n_s-1)\;,
\end{equation}
where 
\begin{equation}
\gamma=\frac{2(\lambda y^2 - 6)}{\lambda y^2} \left[1-\frac{\lambda y^2(\alpha-6)+2(\alpha+3)(\alpha+6)} {(\lambda y^2-2\alpha-6)^2}\right],
\end{equation}
for some selected values of $\alpha$. Note that, in the limit $\alpha \rightarrow 0$ or, equivalently, $\gamma\rightarrow 2$,  Eq. (\ref{nsr}) reduces to $r_{\alpha=0}=8(1-n_s)$, as predicted by exponential models  \cite{melchi}. Also, and very important, we note from this figure that, for these selected values of the parameter $\alpha$, the inflationary scenario discussed in this  paper seems to be in agreement with current observational data from CMB and LSS measurements. As a first example, let us take the tensor fraction $r$ to be negligible. In this case, the analyses involving WMAP3 plus SDSS and WMAP3 plus 2dFGRS data provide, respectively, $n_s = 0.980 \pm 0.020$ and $n_s = 0.956 \pm 0.020$ (68.3\% c.l.), which are clearly in agreement with the model predictions (at $2\sigma$ level) shown in Fig. (3), i.e., $n_s \simeq 1.03$ ($N = 47$), $n_s \simeq 1.01$ ($N = 54$), and $n_s \simeq 0.97$ ($N = 61$). Similar conclusions can also be obtained by considering $r \neq 0$. In this case, the current data from WMAP3 plus SDSS provides a tensor fraction $r < 0.33$ and $n = 0.980 \pm 0.020$, while the model discussed in this paper predicts for this interval of $r$, $n_s \geq 0.95$ ($N = 47$), $n_s \geq 0.92$ ($N = 54$), and $n_s \geq 0.88$ ($N = 61$). From this figure, it is also possible to obtain a scale-invariant spectrum ($n_s = 1$) for values of $r \neq 0$, as discussed in the context of the intermediate inflationary model of Ref. \cite{b1}.

\subsection{Running of the Spectral Index}

The running of the spectral index in the inflationary regime, to lowest order in the {\sl slow-roll} approximation, is given  by \cite{running}
\be
\label{running}
\frac{dn_s}{d\ln k}=-2\xi^2+16\epsilon\eta-24\epsilon^2
\ee
where $\epsilon$ and $\eta$ are, respectively, the first and the second {\sl slow-roll} parameters, defined in Eqs. (\ref{epsilon}) and (\ref{eta}). 
Here, $\xi^2$ is the third {\sl slow-roll} parameter, which is related with the third derivative of the potential by 
\begin{widetext}
\bea
\label{xi}
\xi^2 &=& \frac{m_{\rm pl}^2}{64\pi^2}\frac{V'''V'}{V^2} = \lambda\frac{[6\alpha(2\alpha+3)-3(\alpha+2)\lambda y^2+\lambda^2y^4]
[\lambda y^2-2(\alpha+3)]y^2}{(6-\lambda y^2)^2}\;.
\eea
\end{widetext}
Note that in the limit $\alpha \rightarrow 0$, the $\xi$ parameter reduces to $\xi^2_{\alpha\rightarrow0}=\lambda^2$ and, as expected for usual exponential potentials, the running, expressed by Eq. (\ref{running}), vanishes.

This and other features of the inflationary scenario discussed in this paper are shown in Figs. (4a) and (4b). In Fig. (4a)  the $\alpha - dn_s/d\ln k$ plane is displayed for values of the number of {\sl e-folds} lying in the interval $N = 54 \pm 7$. Note that, differentlty from other models discussed in the literature (see, e.g. \cite{beta}), this scenario predicts only positive values for the running of the spectral index, which seems to be in full agreement with the WMAP3 data ($-0.17 \leq dn_s/d\ln k \leq -0.02$ at $95.4\%$ c.l.) but only partially compatible with the joint analysis involving WMAP3 and SDSS data ($-0.13\leq dn_s/d\ln k \leq 0.007$ at $95.4\%$) of Ref.~\cite{melchi}. In Fig. (4b) we show the $dn_s/d\ln k-r$ plane for $\alpha = 0.1, 0.09$ and $0.08$. Here, the shadowed region corresponds to the 95.4\% limit on the ratio of tensor-to-scalar pertubations, i.e., $r < 0.38$ \cite{melchi}. As can be seen from this Panel, for two out of the three combinations of the pair $\alpha - N$, the model predictions agree reasonably well with the current bounds from CMB and LSS data.

\section{Final remarks}

Primordial inflation \cite{inflation1} constitutes one of the best and most successful examples of physics at the interface between particle physics and cosmology, with  tremendous consequences on our view and understanding of the observable Universe (see, e.g., \cite{revInf,book,lyth1} for review). Besides being the current favorite paradigm for explaining both the causal origin of structure formation and the Cosmic Microwave Background (CMB) anisotropies, an inflationary epoch in the very early Universe also provides a natural explanation of why the Universe is nearly flat ($\Omega_k \simeq 0$), as evidenced by the combination of the position of the first acoustic peak of the CMB power spectrum and the current value of the Hubble parameter \cite{wmap3}.

In this work, we have discussed cosmological implications of the single, minimally-coupled scalar field model recently proposed in Ref. \cite{prl}, whose evolution is described by an exponential potential $V(\Phi)$ that has a quadratic dependence on the field $\Phi$ in addition to the standard linear term. As discussed in Sec. II, this potential fully reproduces the Ratra-Peebles inflation studied in Ref. \cite{exp} in the limit of the dimensionless parameter $\alpha \rightarrow 0$. We have calculated the main observable quantities in the {\sl slow-roll} regime and shown that, even for values of the number of {\sl e-folds} in the restrictive interval $N = 54 \pm 7$ \cite{lyth}, the predictions of the model for values of $\alpha \neq 0$ seem to be in good agreement with current bounds on these parameters from CMB
and LSS observations, as given in Refs. \cite{melchi,tegmark}. Similarly to the intermediate inflationary scenario discussed in Ref. \cite{b1}, it is also possible to obtain a scale-invariant spectrum $(n_s=1)$ for vanishing values of the tensor-to-scalar ratio $r$. For values of $r \simeq 0$ or, equivalently,  $n_s \simeq 1$, we have found that the theoretical prediction for the running of the spectral index approaches to zero from negative values, which is compatible with current observations from CMB data, i.e., $-0.17 \leq dn_s/d\ln k \leq -0.02$ (at $95.4\%$ c.l.)~\cite{wmap3}.

This work is partially supported by the Conselho Nacional de Desenvolvimento Cient\'{\i}fico e Tecnol\'{o}gico (CNPq - Brazil). JSA is also supported by FAPERJ No. E-26/171.251/2004 and JASL by FAPESP No. 04/13668-0.

\end{document}